\documentclass[11pt]{article}

\usepackage{amsmath}
\usepackage{latexsym}
\usepackage{cite}
\usepackage{graphicx}
\usepackage[small,bf,hang]{caption}

\newcommand{\fdual}{\tilde{F}}

\newcommand{\medsp}{\\[0.7ex]}

\newcommand{\veff}{V_{\mbox{\tiny eff}}}

\newcommand{\Ltext}[1]{\ensuremath{\itindex{\mathcal{L}}{#1}}}

\newcommand{\diff}[1]{\mbox{d}#1}

\newcommand{\half}[1]{\ensuremath{\frac{#1}{2}}}
\newcommand{\intd}[1]{\int \!\! #1 \;}
\newcommand{\inv}[1]{\ensuremath{\frac{1}{#1}}}

\newcommand{\metrtilde}[1][]{\Tilde{g}_{\varphi \bar{\varphi} #1}}

\newcommand{\pathd}[1]{\mathcal{D} #1 \;}

\newcommand{\itindex}[2]{\ensuremath{#1_{\mbox{\scriptsize{\itshape #2}}}}}

\newcommand{\varfrac}[2][]{\frac{\delta #1}{\delta #2}}

\DeclareMathOperator{\tr}{Tr}
\DeclareMathOperator{\hc}{h.c.}

\title{Dynamics of Glue-Balls in $N=1$ SYM Theory}

\author{L. Bergamin\\Technische Universit\"{a}t Wien\\bergamin@tph.tuwien.ac.at}

\date{October 6, 2003}

\begin{document}
\maketitle
\begin{abstract}
The extension of the Veneziano-Yankielowicz effective Lagrangian with terms
including covariant derivatives is discussed. This extension is important to
understand glue-ball dynamics of the theory. Though the superpotential remains
unchanged, the physical spectrum exhibits completely new properties.
\end{abstract}

\section{Introduction}
The low energy effective action of $N=1$ SYM theory is written in terms of a
chiral effective field $S = \varphi + \theta \psi + \theta^2 F$, which may be
defined from the local source extension of the SYM action
\cite{veneziano82,Shore:1983kh,burgess95,bergamin01}
\begin{align}
\label{eq:Sdef}
  S &\propto \varfrac{J} W[J, \bar{J}]\ , &
  e^{ i W[J, \bar{J}]} &= \intd{\pathd{V}} e^{i
  \intd{d^4x d^2 \theta} (J + \tau_0) \tr W^\alpha W_\alpha + \hc
  }\ .
\end{align}
With appropriate normalization $S$ is
equivalent to the anomaly multiplet $\bar{D}^{\dot{\alpha}} V_{\alpha \dot{\alpha}} =
D_\alpha S$. $J(x)$ is the chiral source
multiplet, with respect to which a Legendre transformation can be defined
\cite{burgess95,bergamin01}.
The resulting effective action is formulated in terms of the gluino condensate $\varphi
\propto \tr \lambda \lambda$, the glue-ball operators $F \propto \tr F_{\mu
  \nu} F^{\mu \nu} + i \tr F_{\mu \nu} \fdual^{\mu \nu}$ and a spinor $\psi
\propto (\sigma^{\mu \nu}\lambda)_\alpha F_{\mu \nu}$. An effective Lagrangian
in terms of this effective field $S$ has the form \cite{veneziano82,Shore:1983kh}
\begin{equation}
  \label{eq:vylag}
  \Ltext{eff} = \intd{\diff{^4 \theta}} K(S, \bar{S}) - \Bigl( \intd{\diff{^2
  \theta}} S(\log \frac{S}{\Lambda^3} - 1) + \hc \Bigr)\ .
\end{equation}
The correct anomaly structure is realized by the superpotential and thus $K(S,\bar{S})$
is invariant under all symmetries.
In ref.\ \cite{veneziano82} the explicit ansatz $K = k
(\bar{S}S)^{1/3}$ had been made, which leads to chiral symmetry breaking due to $\langle S \rangle =
\Lambda^3$, but supersymmetry is not broken as  $\varphi$
and $\psi$ acquire the same mass $m = \Lambda / k$.
\section{Glue-balls and constraint K\"{a}hler geometry}
Though the spectrum found in ref.\ \cite{veneziano82} does not include any
glue-balls, such fields do appear in $F$. However, they drop out in the analysis of
\cite{veneziano82}, as $F$ is treated as an auxiliary field. Indeed, the
highest component of a chiral
superfield is auxiliary in standard SUSY non-linear\ $\sigma$-models, i.e.\ there appear no derivatives
acting onto this field and moreover its potential is not bounded from below,
but from above. In case of the Veneziano-Yankielowicz Lagrangian the part
depending on the auxiliary field reads
\begin{equation}
  \label{eq:auxlag}
  \Ltext{aux} = k (\bar{\varphi} \varphi)^{- \frac{2}{3}} \bar{F} F + \Bigl(
  \inv{3} \varphi^{-\frac{2}{3}} \bar{\varphi}^{-\frac{5}{3}} F \bar{\psi}
  \bar{\psi} - F \log\frac{\varphi}{\Lambda^3} + \hc \Bigr)\ ,
\end{equation}
and the supersymmetric spectrum is obtained, \emph{if and only if} $F$ is
eliminated by the algebraic equations of motion that follow from
\eqref{eq:auxlag}. This leads to the
unsatisfactory result that glue-balls cannot be introduced in a
straightforward way (cf.\ also \cite{farrar98}) which, in addition, contradicts
available lattice-data \cite{Peetz:2002sr}.

However, in the special case of $N=1$ SYM the elimination of $F$ is not consistent:
If $F$ is eliminated from  \eqref{eq:auxlag}, this implies that the theory
must be ultra-local in the field $F$ \emph{exactly}, i.e.\ even corrections to
the effective Lagrangian which are not included in \eqref{eq:vylag} are not
allowed to change the non-dynamical character of $F$. If this field would be
related to the fundamental auxiliary field, this restriction would be
obvious. But in $N=1$ SYM the situation is different: $S$ is the effective
field from a composite operator and $F$ is not at all related to the
fundamental auxiliary field $D$. As a consequence, the restriction of
ultra-locality on $F$ leads to an untenable constraint on the \emph{physical} glue-ball
operators (for details we refer to
\cite{bergamin01,bergamin02:1,Bergamin:2003ub}).

As shown in ref.\ \cite{Shore:1983kh}, the effective Lagrangian of
\cite{veneziano82} is not the most general expression compatible with all the
symmetries, but the constant $k$ may be generalized to a function $k(\frac{S^{1/3}}{\bar{D}^2 \bar{S}^{1/2}},
  \frac{\bar{S}^{1/3}}{D^2 S^{1/2}})$.
This non-holomorphic part automatically produces space-time derivatives onto
the field $F$, which is most easily seen when $K(S,\bar{S})$ is
rewritten in terms of two chiral fields \cite{Bergamin:2003ub}:
\begin{equation}
  \label{eq:constrkahler}
  K(S, \bar{S}) \rightarrow K(\Psi_0, \Psi_1; \bar{\Psi}_0, \bar{\Psi}_1)
\end{equation}
$\Psi_0$ and $\Psi_1$ are not independent, but they must obey the
constraints
\begin{align}
  \Psi_0 &= S^{\frac13} = \varphi^{\frac13} + \inv{3} \varphi^{-\frac23} \theta \psi +
  \inv{3} \theta^2 (\varphi^{-\frac23} F + \inv{3} \varphi^{-\frac53} \psi \psi)\ , \\
  \Psi_1 &= \bar{D}^2 \bar{\Psi}_0 = \inv{3} (\bar{\varphi}^{-\frac23} \bar{F} +
  \inv{3} \bar{\varphi}^{-\frac53} \bar{\psi} \bar{\psi}) - \frac i 3 \theta
  \sigma^\mu \partial_\mu (\bar{\varphi}^{-\frac23} \bar{\psi}) - \theta^2 \Box
  \bar{\varphi}^{\frac13}\ .
\end{align}
As $F$ appears as lowest component of $\bar{\Psi}_1$, the Lagrangian includes
a kinetic term for that field. In contrast to the situation in
\cite{veneziano82}, this is not inconsistent as the potential in $F$ may
include arbitrary powers in that field (instead of a quadratic term only) and
can be chosen to be bounded from below (instead of above). This way the field
$F$ is promoted to a usual physical field. It has been shown in
\cite{bergamin02:1} that there exist consistent models of this type. In \cite{Bergamin:2003ub} these
ideas have been applied to $N=1$ SYM, leading to an effective action of that
theory with dynamical glue-balls as part of the low-energy spectrum. Formally,
the effective potential looks the same as in the case of Veneziano and Yankielowicz:
\begin{equation}
  \label{eq:veff}
  \begin{split}
    \veff &= - \metrtilde F \Bar{F} + \half{1}
    \metrtilde[, \bar{\varphi}] F (\Bar{\psi} \Bar{\psi}) + \half{1}
    \metrtilde[, \varphi] \Bar{F} (\psi \psi) -
    \inv{4} \metrtilde[, \varphi \Bar{\varphi}] (\psi
    \psi) (\Bar{\psi} \Bar{\psi}) \medsp
    &\quad + c \bigl( F \log \frac{\varphi}{\Lambda^3} + \bar{F} \log \frac{\bar{\varphi}}{\bar{\Lambda}^3}
      - \inv{2 \varphi} (\psi \psi)  - \inv{2 \bar{\varphi}} (\bar{\psi}
    \bar{\psi})  \bigr)
  \end{split}
\end{equation}
However, in contrast to \cite{veneziano82} the K\"{a}hler
``metric''\footnote{This quantity is not equivalent to the true K\"{a}hler
  metric of the manifold spanned by $\Psi_0$ and $\Psi_1$, cf.\ \cite{Bergamin:2003ub}.}
is a function of $\varphi$ \emph{and} $F$, $\metrtilde(\varphi, F;
\bar{\varphi}, \bar{F})$. From eq.\ \eqref{eq:veff} the
consistent vacua can be derived, for explicit expressions we
refer to \cite{Bergamin:2003ub}. The most important properties of the
Lagrangian \eqref{eq:vylag} with \eqref{eq:constrkahler} are:

The effective potential is minimized with respect to \emph{all} fields
    $\varphi$, $\psi$ and $F$. Consequently, the dominant contributions that
    stabilize the potential must stem from the
    K\"{a}hler part, not from the superpotential: The superpotential is a
    holomorphic function in its fields and therefore its scalar part must have
    unstable directions. In the present context there exists no mechanism to transform
    these instabilities into stable but non-holomorphic terms.

Though the model has the same superpotential as the
Lagrangian of ref.\ \cite{veneziano82} its spectrum is completely different: Chiral symmetry breaks by a vacuum expectation value (vev) of $\varphi
        \propto \Lambda^3$, but this mechanism is more complicated than in
        \cite{veneziano82}. Any stable ground-state must have non-vanishing vev of
        $F$. But
        $\langle F\rangle$ is the order parameter of supersymmetry breaking and thus this
        symmetry is broken as well\footnote{The author of ref.\
    \cite{Shore:1983kh} concluded that this model cannot have a stable \emph{supersymmetric}
    ground-state. This is in agreement with our results, as the model breaks down as $F \rightarrow 0$.}. $\psi$ is a massless spinor, the Goldstino.

The supersymmetry breaking scenario is of essentially non-perturbative
    nature\footnote{The importance of such a breaking mechanism has been
      pointed out in \cite{bergamin01} already, but a concrete description was not yet
    found therein.}: it is not compatible with perturbative non-renormalization
    theorems, as the value of $\veff$ in its minimum and the
    vev of ${T^\mu}_\mu$ are no longer
    equivalent. In particular, the former can be negative, while the latter is
    positively semi-definite due to the underlying current-algebra
    relations. To our knowledge this is the first model, where this type of
    supersymmetry breaking has found a concrete description (cf.\
    \cite{bergamin02:1,Bergamin:2003ub} for details).

Any ground state with $\langle \metrtilde \rangle \neq 0$ can be equipped with stable
    dynamics for $p^2 < |\Lambda|^2$. In the construction of concrete kinetic
    terms it is important to realize that \eqref{eq:constrkahler} may include
    expressions with explicit space-time derivatives. Again this is possible as
    $F$ is not interpreted as an auxiliary field. 

In summary, the Lagrangian of ref.\ \cite{Bergamin:2003ub} is the most
general one, which can be formulated in terms of the effective field
$S$. Consistent ground-states can be found together with broken supersymmetry
only. It would be interesting to compare these results with a \emph{different}
action, which has supersymmetric ground-states. But the "pi\`{e}ce de
r\'{e}sistance'' for such an action is the fact, that it cannot start from the
effective field $S$. 
\subsection*{Acknowledgements}
The author would like to thank U.~Ellwanger, J.-P.~Derendinger, E.~Kraus,
P.~Minkowski, Ch.~Rupp and E.~Scheidegger for interesting discussions. This
work has been supported by the Austrian Science Foundation (FWF) project P-16030-N08.

\end{document}